\documentclass[12pt]{iopart}
\usepackage{iopams}
\usepackage{graphicx}
\usepackage{dcolumn}
\usepackage{bm}
\usepackage{hyperref}

\begin{document}

\title{Experimental demonstration of an atomtronic battery}

\author{Seth C. Caliga$^1$, Cameron J. E. Straatsma$^2$, and Dana Z. Anderson$^1$}
\address{$^1$ JILA and Department of Physics, University of Colorado and National Institute of Standards and Technology, Boulder, Colorado 80309-0440}
\address{$^2$ JILA and Department of Electrical, Computer, and Energy Engineering, University of Colorado, Boulder, Colorado 80309-0440}

\begin{abstract}
Operation of an atomtronic battery is demonstrated where a finite-temperature Bose-Einstein condensate stored in one half of a double-well potential is coupled to an initially empty load well that is impedance matched by a resonant terminator beam. The atom number and temperature of the condensate are monitored during the discharge cycle, and are used to calculate fundamental properties of the battery.  The discharge behavior is analyzed according to a Th\'{e}venin equivalent circuit that contains a finite internal resistance to account for dissipation in the battery. Battery performance at multiple discharge rates is characterized by the peak power output, and the current and energy capacities of the system.
\end{abstract}
\maketitle

\section{\label{sec:Intro}Introduction\protect}
Fundamental to all classical circuits, electronic and atomtronic alike, is the need for a source of power to drive circuit operation. Within atomtronic systems, chemical potential and atom current play the roles of electric potential and electron current in electronics, respectively. Because they operate in the ultracold regime, temperature plays a more fundamental role in atomtronics than it does in electronics. Current in atomtronic circuits can be driven by thermodynamic gradients, i.e. spatially varying chemical potential and/or temperature. Indeed, atomic systems containing interconnected reservoirs initialized with temperature and chemical potential gradients between the wells have been used to demonstrate behavior analogous to the thermoelectric effect~\cite{2013:Brantut} and that of a discharging RLC circuit~\cite{2013:Lee}. In the quantum domain current can exist even in thermal equilibrium. Atomtronic analogues of superconducting circuits study the behavior of a quantum gas flowing through a weak link~\cite{2014:Eckel, 2014:Jendrzejewski} or tunnel junction~\cite{2006:Gati, 2007:Gati}, where current flow is determined by the phase of the gas. Though an ideal quantum circuit operates without power input, in reality useful atomtronic circuits, classical or not, will require a power source. In this work we consider a self-contained power source, which we refer to simply as ``a battery".

\begin{figure}[t]
\includegraphics[width=8.6cm]{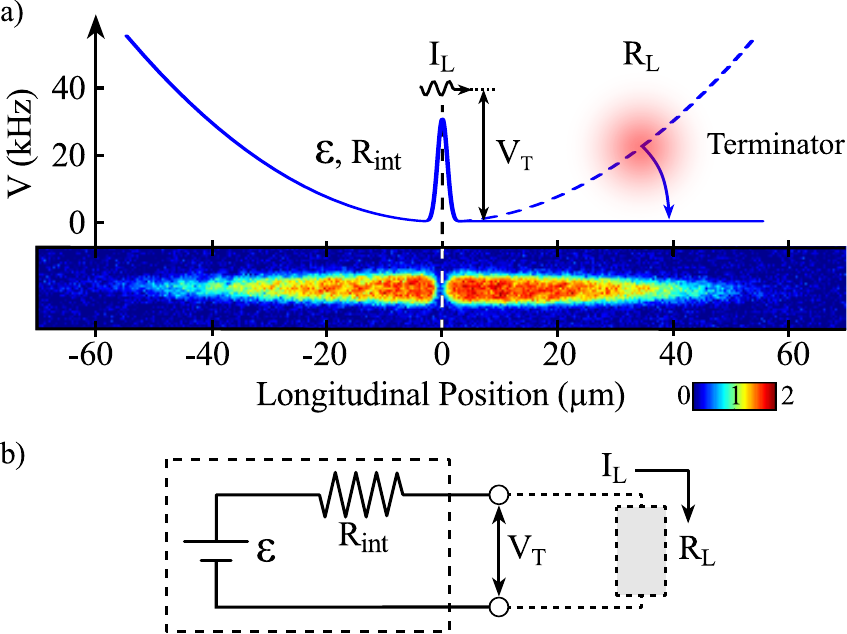}
\centering
\caption{\label{fig:Circuit} 
a) Schematic of the double-well atomtronic battery system.  The top panel shows the longitudinal potential energy landscape.  A resonant ``terminator" beam removes atoms that flow into the load well from the system. The bottom panel shows a false color \emph{in situ} absorption image of atoms occupying both wells of the potential. b) Th\'{e}venin equivalent circuit representation of the atomtronic battery.  Analogous circuit properties are labeled in (a) and (b): electromotive force $\varepsilon$, internal resistance $R_\mathrm{int}$, atom current $I_\mathrm{L}$, terminal voltage $V_\mathrm{T}$, and load resistance $R_\mathrm{L}$. The dashed portion of the circuit represents the terminator beam as a lumped element that is always impedance matched to the battery internal resistance.}  
\end{figure}

Atomtronic batteries have been proposed in open quantum systems comprised of reservoirs with different chemical potentials connected by a lattice~\cite{2007:Seaman, 2010:Pepino}, as well as a mesoscopic system containing a finite-temperature Bose-Einstein condensate (BEC) that sources condensed atoms~\cite{2013:Zozulya}. In this paper, the double-well system shown in Fig.~\ref{fig:Circuit}(a) is used to quantify fundamental characteristics of an atomtronic battery including the effective terminal voltage, peak power output, and the current and energy capacities.  The system consists of a finite temperature BEC with chemical potential $\mu$ and temperature $T$ prepared in one of the two wells while the other is initially empty.  This configuration establishes non-equilibrium thermal and chemical potential gradients that drive current across the central, repulsive barrier and into the ``load" well. Here, we explore the manner in which the stored energy of the reservoir acts as a supply of atom current to the load.

The battery, serving in its intended role, will typically supply atom current to a load consisting of a collection of circuit devices such as transistors, filtering elements, and so on. In contrast to an electrical battery driving a circuit operating at low frequency, however, atom coherence properties play a fundamental role in the behavior of the battery-load system. That is, as a source of current, the atomtronic battery is akin to a microwave power source for which the impedance of both source and load act together to determine system behavior. In this paper, the load well is illuminated with resonant laser light, which, by removing atoms from the well, eliminates possible reflections of atoms back towards the reservoir. By eliminating such reflections, our load is effectively impedance matched to the battery. Generally speaking, impedance matching provides maximum power delivery from the battery.

It may be useful to comment that at first glance the double-well system depicted in Fig.~\ref{fig:Circuit}(a) resembles experimental arrangements used to implement atom lasers~\cite{1999:Hagley,1999:Bloch,2006:Guerin,2008:Couvert}. In particular, atoms are selectively out coupled from the left-hand reservoir into a magnetic guide where they propagate ballistically. Atom lasers, however, utilize coupling with a Bose-Einstein condensate to provide, ideally, a source of fully coherent atoms. In our case, the degree of coherence of the out-coupled atoms is determined by the reservoir temperature, while their mean kinetic energy is determined primarily by the output barrier height. In addition, atom lasers typically utilize weak coupling to the condensate, thereby minimizing back-action of the output on the reservoir. By contrast, we are very much interested in the evolution of the atom reservoir and show the extent to which it can be described in the context of a conventional electronic battery.

Associated with an electronic battery is an internal resistance that dissipates energy as the battery supplies current to a load. The presence of this internal resistance is fundamental to the action of a battery unlike, say, a capacitor, which can in principle supply current to a load without associated dissipation internal to the capacitor. Similarly, an atomtronic battery must likewise exhibit an internal resistance; resistance that induces dissipation that can be seen as a consequence of an open quantum system. An analytical treatment of internal resistance in our double-well system is given in Ref.~\cite{2013:Zozulya}. As is true for the electronic battery, internal resistance determines the maximum power that can be delivered to the load, determines the amount of heat generated internal to the battery, and its energy dissipation causes noise to be injected into the circuit. Operating in the ultracold regime where quantum coherence properties are of interest, heating and noise play a central role (see for example Ref.~\cite{2007:Gati}). In this study of an atomtronic battery, rather than explicitly analyzing the microscopic physical processes associated with dissipation, dissipative effects are cast in terms of an effective internal resistance within the Th\'{e}venin equivalent circuit shown in Fig.~\ref{fig:Circuit}(b).

Having in mind the picture of a ``AA" battery as the prototype, let us here make some conceptual remarks. A battery has two terminals: a ``+" and ``-". If we identify the plus terminal as a source of current, then the second, minus, terminal serves two purposes. On the one hand it defines a voltage reference (often defined as zero volts) against which a potential difference is determined, and on the other hand it serves as a current return, which effectively maintains charge neutrality for the circuit as a whole. In our case, the reference potential is somewhat arbitrary, but to be concrete one might choose as the physical plus terminal location to be just to the left of the barrier and the minus terminal just to the right of the barrier of Fig.~\ref{fig:Circuit}(a). The difference in chemical potential between these two terminals is the battery potential. However, unlike the electronic case, the minus terminal need not serve as a current return.

The remainder of this paper is organized as follows:  In Sec.~\ref{sec:Expt}, the experimental system is described along with the initialization and operation of the battery.  Details regarding the evolution of the reservoir ensemble during the battery discharge cycle are also given.  Section~\ref{sec:Model} introduces the model used to track the decay of both atom number and energy stored within the battery as it is discharged.  The time evolution of the thermodynamic variables of the reservoir well reveals information regarding the dissipative effects associated with current flow and power output.  Results from the experiment and model are used in Sec.~\ref{sec:Circuit} to analyze battery discharge behavior in the context of the Th\'{e}venin equivalent circuit shown in Fig.~\ref{fig:Circuit}(b). From this data it is possible to quantify the internal resistance and terminal voltage of the battery.  The performance of the atomtronic battery, which includes the current and energy capacities, as well as the peak power output, is analyzed in Sec.~\ref{sec:Performance}. Finally, Sec.~\ref{sec:Conclusion} provides concluding remarks. 

\section{\label{sec:Expt}Experimental System \& Battery Operation\protect}
The atomtronic battery is realized in an atom-chip-based apparatus that is described in detail elsewhere~\cite{2013:Salim, 2016:NJP}. Overall harmonic confinement of $^{87}$Rb atoms is provided by magnetic fields generated by current flowing through conductors on the atom chip and external bias coils.  The resulting cigar shaped trap, with frequencies $(\nu_{x},\nu_{\perp}) = (67, 1500)$ Hz, is located $\sim150~\mu$m below an optically transparent region of the atom chip.  To create the double-well structure of the battery shown in Fig.~\ref{fig:Circuit}(a), a repulsive optical barrier of 760 nm light is projected through the chip window onto the magnetic potential. The optical potential is patterned using an acousto-optic deflector (AOD) and has longitudinal and transverse full widths at 1/e peak intensity of $2.1~\mu$m and $16~\mu$m, respectively. 

\begin{figure}[t]
\includegraphics[width=\columnwidth]{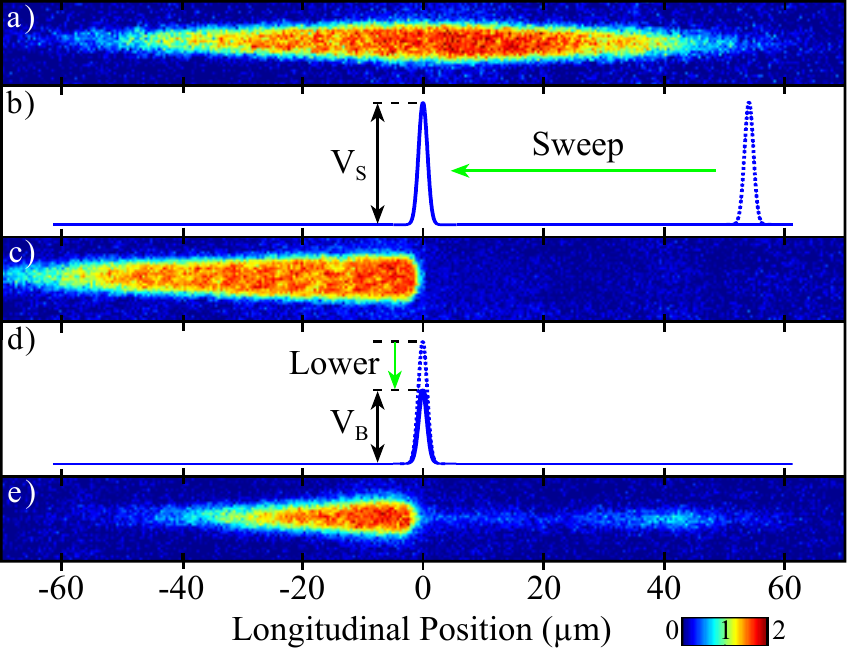}
\centering
\caption{\label{fig:Batt_Init} 
Preparation and discharge of the atomtronic battery. a) False color \emph{in situ} image of the atoms in the bare magnetic potential. b) Longitudinal profile of the repulsive barrier with height $V_{S} = 100$ kHz while loading the reservoir well. c)  Image of the initial state of the battery with the barrier height held at $V_{S}$. d) Reducing the barrier height to $V_{B}$  couples the battery to the test load. e) Image of the battery output 30 ms after the barrier has been lowered to $V_{B} = 30$ kHz.}  
\end{figure}

Initial state preparation of the battery is shown in Fig.~\ref{fig:Batt_Init}. Atoms are loaded into the magnetic chip trap and cooled below the critical temperature $T_{c}$ via forced radio-frequency (RF) evaporation. The repulsive barrier is turned on at the edge of the longitudinal spatial extent of the cloud with a height $V_{S}= 100$ kHz $\gg T$~\footnote{Throughout this paper, the unit convention of Ref.~\cite{2013:Zozulya} is followed, and both energies and particle currents are reported in units of Hz.}. It is then swept adiabatically to its final location at magnetic trap center, as shown in Fig.~\ref{fig:Batt_Init}(b). Dynamic control over the position and height of the barrier is achieved by adjusting the power and frequency of the RF signal driving the AOD.  In this way, all of the atoms are loaded into the reservoir well, and the battery is considered charged.  By lowering the barrier height to $V_{B}$, as shown in Fig.~\ref{fig:Batt_Init}(d), the battery is connected to the load well, and current is allowed to flow from the reservoir well.

In order to prevent the longitudinal harmonic confinement of the load well from reflecting the battery output current back into the reservoir well, atoms that enter the initially empty well are removed from the system.  This is accomplished by illuminating the load well with a resonant light beam, labelled ``Terminator" in Fig.~\ref{fig:Circuit}(a), that optically pumps atoms into untrapped $m_F$ states.  To ensure atoms are only removed from the load well, the terminator beam has a full width at 1/e of $16~\mu$m and is displaced $35~\mu$m from trap center into the empty well.  Thus, the terminator beam provides an effect analogous to impedance matched termination in RF electronics. 

In-trap images are used to measure the spatial distribution of the atoms in the load well and calibrate the height of the repulsive barrier.  By extinguishing the terminator beam $\sim 4$ ms prior to imaging, atoms emitted into the load well are allowed to interact with the magnetic potential for a time equal to roughly one quarter of the longitudinal trap period.  As a result, the images serve as a rudimentary spectrum analyzer, allowing the longitudinal momentum of the atom current to be extracted from the spatial profile of the atoms at their classical turning point. Figure~\ref{fig:Batt_Init}(e) shows the battery output current and reservoir imaged using this technique.

The battery is discharged across barriers ranging in height from $V_{B} = 20\--70$ kHz, in 10 kHz increments, in order to characterize its behavior at different output currents. Bimodal fits to the spatial density profile of the reservoir atoms in time-of-flight are used to calculate the number of thermal, $N_\mathrm{th}$, and condensed, $N_\mathrm{c}$, atoms remaining in the reservoir, as well as the ensemble temperature, at different times during the discharge. 

\begin{figure}[t]
\includegraphics[width=8.6cm]{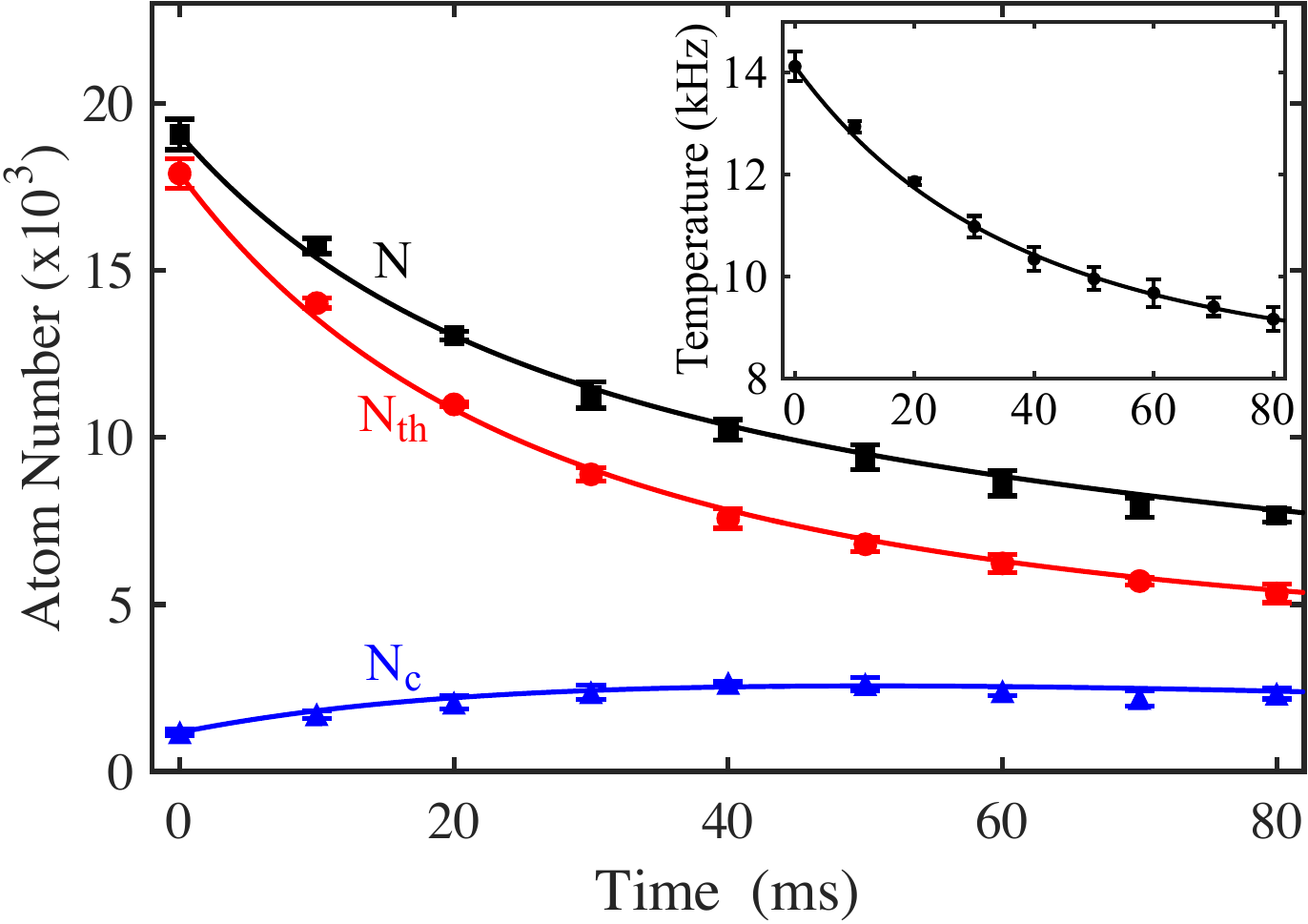}
\centering
\caption{\label{fig:ANum50} 
Time evolution of the total (black squares), thermal (red circles), and condensed (blue triangles) atom number in the reservoir for $V_{B} = 50$ kHz.  The temperature of the reservoir ensemble is shown in the inset. Error bars represent the standard error of the mean for five experimental realizations. Solid lines show the results from numerical calculation of the reservoir population.}  
\end{figure}

Figure~\ref{fig:ANum50} shows a representative data set for the time evolution of the reservoir ensemble during battery discharge with $V_{B} = 50$ kHz. 
In this data set, the total initial reservoir population is $N=19.07(93)\times10^3$, with a condensed population of $N_c=1170(84)$.  Approximating the reservoir as a half-harmonic well with $\nu_{x,R}\approx 2\nu_{x} = 134$ Hz, the chemical potential of the reservoir ensemble is calculated in the Thomas-Fermi limit. Validity of the half-harmonic approximation for calculation of the chemical potential is confirmed to within $<5\%$ error for the experimental parameters by GPE simulation.  Therefore, the initial reservoir population corresponds to a chemical potential and temperature of $\mu = 2.86(8)$ kHz and $T = 14.13(50)$ kHz, respectively.  These potentials drive an initial atom current of $I = 334(22)$ kHz. As atoms flow from the reservoir, across the repulsive barrier, and into the load well, the total reservoir atom number decreases along with the temperature. Decreasing number and energy within the reservoir represent power supplied to the load. Notably, the battery reservoir cools as it discharges since thermal atoms are coupled to the load well, which is a process akin to evaporative cooling of a dilute gas and radiative cooling of a blackbody. After approximately $30$ ms, a nearly constant chemical potential, $\mu = 3.86(16)$ kHz, is reached and then sustained for the remaining $\sim 50$ ms. By $80$ ms the temperature has decreased to $T = 9.16(23)$ kHz, as shown in the inset of Fig.~\ref{fig:ANum50}, and the current to $I = 60.8(7.7)$ kHz. Battery discharge with other barrier heights exhibit similar behavior where lower $V_B$ results in a more rapid depletion of both the reservoir atom number and temperature. This behavior is discussed further in the following sections.
 
\section{\label{sec:Model}Discharge Model\protect\\}
As the battery discharges, only atoms in the high energy tail of the distribution can contribute to the output current (i.e., the effects of tunneling are negligible).  However, the reservoir is in the hydrodynamic regime along the longitudinal dimension of the trap ensuring rapid rethermalization such that the reservoir ensemble can be assumed to maintain a state of dynamical equilibrium. As such, we model the discharge behavior of the atomtronic battery using methods of previous works that study evaporation and trap loading processes~\cite{2003:Roos, 1996:Walraven}.  The evolution of atom number and energy stored in the reservoir are described by particle and energy currents:
\begin{eqnarray}
\frac{dN}{dt} & = & -\gamma N_{th} f\!\left(\eta,\mu,T \right), \label{eqn:rate_tot}\\ 
\frac{dE}{dt} & = & \frac{dN}{dt} \left(\eta\!+\!\kappa \right)T+\beta C.\label{eqn:rate_enrg} 
\end{eqnarray}
The magnitude of the particle current is governed by the probability that an atom has sufficient energy to traverse the barrier, $f(\eta,\mu,T) = \exp{\left(\mu/T-\eta\right)}$, where $\eta = V_{B}/T$ is the truncation parameter. This probability is scaled by the quantity $\gamma N_{th}$, which reflects the rate at which atoms interact with the barrier. Here, the binary collision rate is given by $\gamma = 32 \zeta(3/2)m(\pi a_{s}T)^2/h$~\cite{2002:Pethick}, where $\zeta$ is the Riemann-Zeta function, $m$ is the atomic mass, $a_s$ is the s-wave scattering length, and $h$ is the Planck constant.  The energy current is given by the product of the particle current and the average energy removed per particle given by $\left(\eta\!+\!\kappa \right)T$.  The factor $\kappa$ indicates the average energy each particle removes from the reservoir in excess of the barrier height energy.   The value of $\kappa$ is of order unity and depends on the dimensionality of the momentum distribution of the ensemble relative to the allowed escape trajectories~\cite{2003:Roos, 1996:Walraven}.  For the potential studied here, $\kappa \sim 3$, whereas in isotropic evaporation (e.g., forced RF evaporation) $\kappa \sim 1$~\cite{1996:Ketterle}.

To account for additional heating or cooling effects external to the flow of atoms the rate $\beta C$ is included. This term does not affect the total atom number, but transfers energy to the reservoir according to the heat capacity $C$ and external heating or cooling rate $\beta$.  In the context of this demonstration only heating due to technical trap noise is included~\footnote{The heating rate due to technical noise contributes $\sim 1.6$ kHz of thermal energy to the reservoir over the measured 80 ms discharge period.}; however, this term can be used to describe less trivial mechanisms such as heat generation associated with irreversible processes during circuit operation~\cite{1961:Landauer}. The derivation of Eqs.~(\ref{eqn:rate_tot}) and~(\ref{eqn:rate_enrg}) accounts for interactions between the condensed and thermal portions of the gas, and for our trap geometry are valid for temperatures in the range $T_c > T > T_0$, where $T_0$ is the interaction energy per particle~\cite{2002:Pethick}.  

We determine $\kappa$ from experimental data across a range of truncation parameters using the relation
\begin{equation}
\label{eqn:kappa}
\kappa = \left(\frac{3}{2}+\delta\right)\left(\alpha+1\right)-\eta,
\end{equation}
where $\delta = 3/2$ for a three-dimensional harmonic oscillator~\cite{1996:Ketterle}. Here, $\alpha$ represents the energy transfer efficiency given by
\begin{equation}
\label{eqn:alpha}
\alpha = \frac{\dot{T}/T}{\dot{N}/N},
\end{equation}
where $\dot{N}$ and $\dot{T}$ are the time derivatives of the number of reservoir atoms and their temperature, respectively, and are extracted from the time evolution of the battery reservoir (shown in Fig.~\ref{fig:ANum50}) using a first-order finite-difference method.   
\begin{figure}[t]
\includegraphics[width=8.6cm]{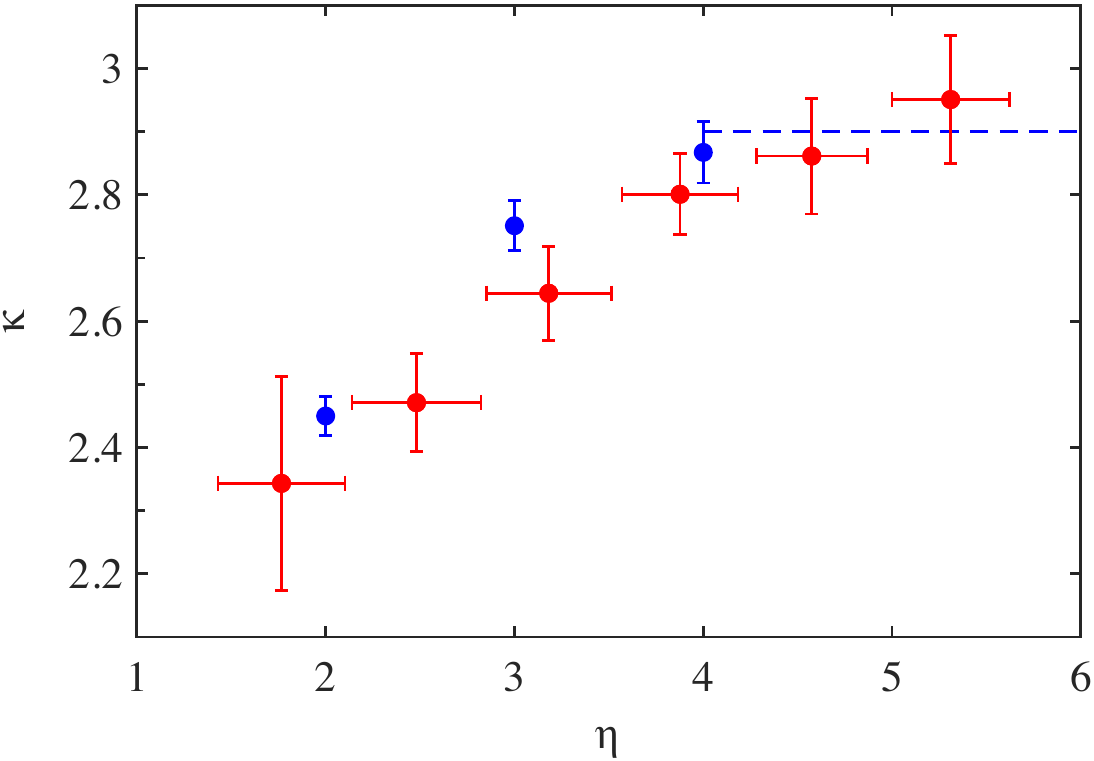}
\centering
\caption{\label{fig:kappa} 
Value of the $\kappa$ parameter from experiment (red circles), DSMC simulations (blue circles), and Ref.~\cite{2003:Roos} (blue dashed line). Horizontal and vertical error bars on the experimental data represent the standard deviation of $\eta$ and $\kappa$ during the time evolution of each data set, respectively. Errors bars on the DSMC results represent the spread in energy of atoms entering the load well.}  
\end{figure}
Experimental values of $\kappa$ for various truncation parameters are plotted in Fig.~\ref{fig:kappa} as red circles.  The horizontal and vertical positions of each point indicate $\bar{\eta}$ and $\bar{\kappa}$, the mean values of $\eta$ and $\kappa$ across the measured discharge time, respectively.  Similarly, error bars indicate the range of $\eta$ and $\kappa$ values during discharge and incorporate statistical uncertainty from five experimental realizations. The horizontal dashed line indicates results of Ref.~\cite{2003:Roos}.  For $\eta<4$ the value of $\kappa$ decreases below the simulated value of $\kappa = 2.9$ as the transverse momentum distribution begins to play a role in the magnitude of the atom current, effectively reducing $\kappa$. This effect is confirmed through simulations using the direct simulation Monte Carlo (DSMC) method~\cite{birdbook}, results of which are plotted in Fig.~\ref{fig:kappa} as blue circles. The error bars represent the spread in energy of atoms entering the load well. For application of the DSMC method, the finite temperature BEC is treated as an ideal gas that evolves in the presence of a static mean-field potential representing a BEC in the Thomas-Fermi limit. We find this approximation to be suitable for the system studied here as only the high energy tail of the distribution is coupled to the load well.

To compare the model to experimental data, the solution to Eqs.~(\ref{eqn:rate_tot}) and~(\ref{eqn:rate_enrg}) is calculated using a fourth order Runge-Kutta method in which parameters of the ensemble (e.g. $N$, $T$, and $C$) are updated at each time step using expressions derived in Ref.~\cite{2002:Pethick}.  Experimental parameters including the mean values of $\kappa$ and $\eta$ from Fig.~\ref{fig:kappa} along with initial reservoir numbers and temperatures are used to initialize the calculation.  Results of the numerical calculation for $N$, $N_{th}$, $N_{c}$, and $T$ of the battery reservoir during discharge across a barrier with height $V_{B} = 50$ kHz are shown as solid lines in Fig.~\ref{fig:ANum50}.  We find the model to be in good agreement with the experimental data.

\section{\label{sec:Circuit}Th\'{e}venin Equivalent Circuit\protect}
We analyze the atomtronic battery using the Th\'{e}venin equivalent circuit of the analogous electronic battery shown in Fig.~\ref{fig:Circuit}(b).  Of particular interest is the role of dissipation in the atomtronic system.  Dissipation in electronic devices manifests in the form of heat generated during the flow of current (i.e., Ohmic heating). Electronic batteries store energy that gives rise to an electromotive force $\varepsilon$, and contain a finite internal resistance $R_{int}$.  When connected to a load with resistance $R_{L}$, current $I_{L} = \varepsilon/\left(R_{int}+R_{L}\right)$ flows through the circuit. Due to the non-zero internal resistance, power is dissipated within the battery equivalent to $P_{int}=I_{L}^2R_{int}$.  These relations are used to study dissipation in the atomtronic system, where $\mu$ replaces $\varepsilon$ and $R_{int}$ represents dissipation in the reservoir.

Th\'{e}venin's theorem is applied to the battery as a way to describe the complex effects of dissipation in the battery reservoir with a single quantity $R_{int}$. Although strictly valid only for linear circuit elements, the Th\'{e}venin equivalent circuit here can be considered as a small-signal equivalent circuit that describes device operation about some quiescent point. As is generally true for AC circuits, the Th\'{e}venin equivalent resistance is frequency dependent, i.e., energy dependent for an atomtronic circuit. Indeed, as shown in~\cite{2013:Zozulya}, the internal resistance can range from negative to positive depending on the energy of the supplied current. In our experiments, the resistance is well approximated by a single value associated with the barrier height since the thermal energy is small compared with that height. Note that we do not include any thermoelectric effects in our analysis as there is no temperature gradient in the portion of the system represented by an equivalent circuit.

The magnitude of the internal resistance in a battery is a useful metric for performance and applications, which will be discussed further in Sec.~\ref{sec:Performance}. For example, an application requiring a large peak output power (i.e., starting a car engine) requires a battery with a low internal resistance. Otherwise, the terminal voltage of the battery drops to zero as soon as the battery is connected to the load. As mentioned in the Introduction, the presence of internal resistance in a battery ultimately determines the maximum power available to drive a circuit, and imposes noise onto the circuit.

As will be shown below, in the atomtronics case, the sign of the internal resistance is indicative of the atoms supplied to the load. In this work, thermal atoms are supplied and the internal resistance is negative. This indicates that the reservoir cools and entropy decreases locally. On the other hand, if condensed atoms were supplied to the load, as in Ref.~\cite{2013:Zozulya}, then the internal resistance would be positive, indicating that the reservoir heats and entropy increases locally. Thus, the presence of internal resistance shows that any supply of power to an atomtronic circuit, whether it be supplying thermal or condensed atoms, will internally have dissipation.

\begin{figure}[tb]
\includegraphics[width=8.6cm]{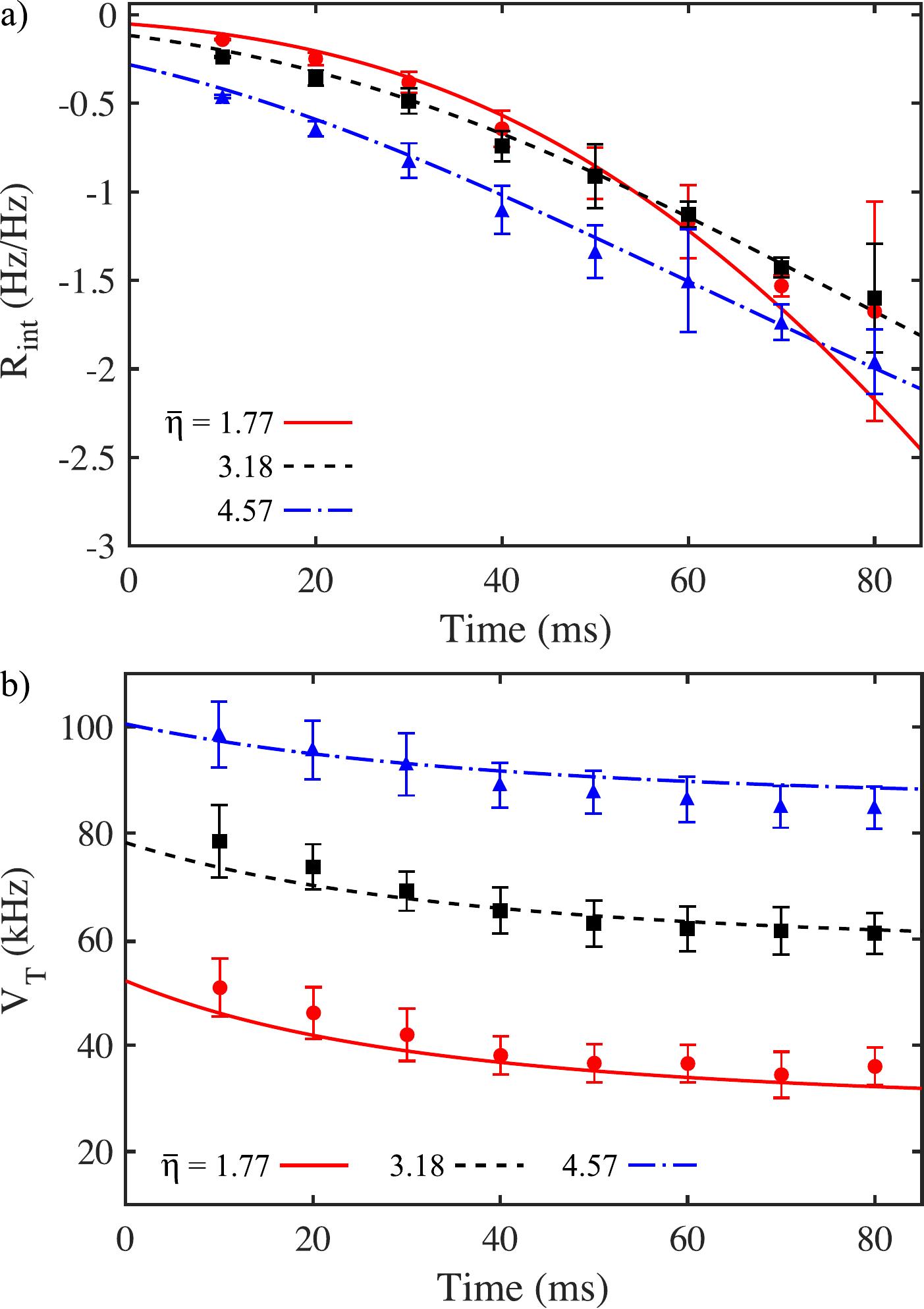}
\centering
\caption{\label{fig:VTRint} 
a) Internal resistance during battery discharge for three barrier heights with mean truncation parameters of $\bar{\eta} = 1.77$ (solid, red), $3.18$ (dashed, black), and $4.57$ (dash-dot, blue). b) Effective terminal voltage during battery discharge corresponding to data in (a). Data points show the corresponding results for $R_{int}$ and $V_{T}$ from the experimental data calculated using Eq.~(\ref{eqn:Rint}) and Eq.~(\ref{eqn:VT}), respectively. Error bars are propagated from the standard error of the mean for $N$, $\mu$, and $T$.}  
\end{figure}

The power dissipated within the atomtronic battery is calculated using the results for the evolution of the reservoir ensemble during battery discharge.  Here, power dissipated within the battery is equivalent to the time rate of change in internal energy of the ensemble within the static reservoir well given by $P_{int} = \dot{E} - \mu \dot{N}$; thus,
\begin{eqnarray}
R_{int} & = & \frac{\dot{E} - \mu \dot{N}}{I^2_{L}} \nonumber \\
            & \simeq & \frac{\mu-\left(\eta+\kappa\right)T}{I_{L}}, \label{eqn:Rint}
\end{eqnarray}
which has units of Hz/Hz. This quantity encapsulates the heating rate or entropy production inside the reservoir. In simplifying the expression for $R_{int}$ to obtain Eq.~(\ref{eqn:Rint}), $\dot{N}$ and $\dot{E}$ have been replaced by Eqs.~(\ref{eqn:rate_tot}) and~(\ref{eqn:rate_enrg}), respectively, and $I_{L} = -\dot{N}$. Additionally, an additive factor of $\beta C/I^2_{L}$ that arises due to technical heating of the reservoir has been omitted, as it is small compared to other contributions to the internal resistance. We find that the internal resistance is negative, corresponding with the observed cooling of the reservoir, and its magnitude increases in time. Figure~\ref{fig:VTRint}(a) plots the internal resistance within the atomtronic battery as it discharges.  Evolution of the internal resistance is shown for three representative barrier heights, denoted by the mean truncation parameter of the barrier, $\bar{\eta}$.  As the battery discharges the increase in $|R_{int}|$ indicates an increased impedance to current flow.  

The terminal voltage of the Th\'{e}venin equivalent circuit is given by
\begin{eqnarray}
V_{T} & = & \mu - I_{L} R_{int} \nonumber \\
          & \simeq & (\eta+\kappa)T, \label{eqn:VT}
\end{eqnarray}
where Eq.~(\ref{eqn:Rint}) has been inserted for $R_{int}$.  Therefore, the terminal voltage is the average energy carried by each atom as it leaves the battery. Note that the potential for atoms leaving the battery to do work on a load is given by the free energy, which requires an estimation of the entropy produced in the out coupling process. We do not investigate this further here. The time-evolution of $V_{T}$ corresponding to the $R_{int}$ data is shown in Fig.~\ref{fig:VTRint}(b). Depletion of stored energy accompanied by increasing $R_{int}$ leads to a decrease in the voltage supplied by the battery, as is the case in electronic batteries.  The terminal voltage of the fully charged atomtronic battery depends on both the height of the barrier and the reservoir ensemble temperature. As the battery discharges, the terminal voltage evolves to first order according to $\dot{V}_{T}\propto \kappa\dot{T}$, which reflects the decrease in the mean energy of the atoms contributing to the load current as the reservoir cools.

\section{\label{sec:Performance}Battery Performance\protect}
\begin{figure}[!bt]
\includegraphics[width=8.6cm]{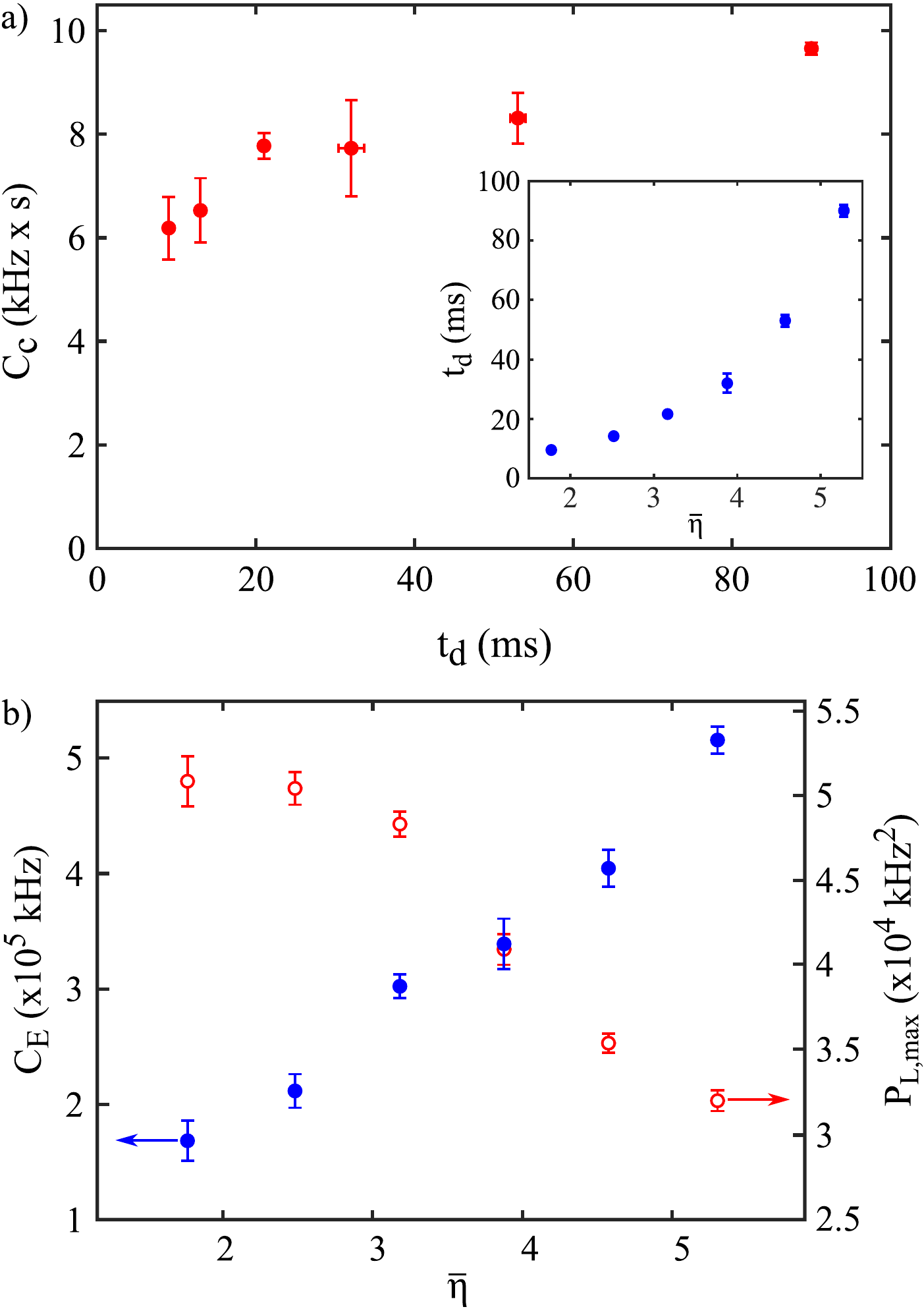}
\centering
\caption{\label{fig:BCc}
a) Current capacity $C_c$, shown here in units of kHz~$\times$~s, is found to increase monotonically with the discharge time, $t_d$, which is shown in the inset.  The discharge time increases exponentially with the mean truncation parameter, which is consistent with Eq.~(\ref{eqn:rate_tot}). b) Energy capacity, $C_E$, (filled blue circles, left axis) increases linearly with increasing barrier height. Peak power output $P_{L,peak}$ (open red circles, right axis) decreases monotonically with increasing barrier height. Error bars are propagated from the standard error of the mean for $N$, $\mu$, and $T$.}  
\end{figure}

Performance of the atomtronic battery is governed by the truncation parameter $\bar{\eta}$ and can be characterized by both the capacity of the battery and the peak power delivered to the impedance matched load~\footnote{Impedance matching implies that $P_{L} = V^{2}_{T}/R_{L}$, where $R_{L}=R_{int}$. Using the calculated values of $R_{int}$ and $V_{T}$, the load resistance is calculated using $R_{L} = R_{int}/(\mu/V_{T}-1)$. For all data sets we find $|R_{L}| = |R_{int}|$, confirming that the load well is impedance matched by the terminator beam.}. Electronic battery capacity is characterized in multiple ways; however, two common metrics are the total current and total energy delivered to the load during the time it takes the terminal voltage to drop below a set depletion voltage~\cite{2011:Zhang}.  Here, the discharge time $t_d$ is defined as the time it takes the terminal voltage to fall below 90\% of its initial value. The current capacity $C_c = \int^{t_d}_0 I_L dt$ of the atomtronic battery is then calculated by integrating the current output from the time the barrier is lowered from $V_{S} \rightarrow V_{B}$ until $t_d$.  Figure~\ref{fig:BCc}(a) plots the current capacity as a function of the discharge time for $\bar{\eta} \approx 1.77 \-- 5.28$.

Current capacity in electronic batteries is often characterized in terms of Peukert's law, which states that battery current capacity obeys the relationship $C_c= I^\rho_L t_d$, where $\rho$ is the Peukert constant~\cite{2008:Jongerden}.  An ideal battery is defined to have $\rho = 1$, while typical lead-acid electronic batteries range from $\rho = 1.1\-- 1.5$, where $\rho > 1$ indicates internal power loss~\cite{2016:Jung}. Therefore, by increasing $t_d$ the current capacity of the battery increases as well. Indeed, the data in Fig.~\ref{fig:BCc}(a) indicates that by increasing the discharge time by roughly an order of magnitude, the current capacity increases by a factor of $1.6(2)$.  From measured values for the current and discharge time, along with the calculated current capacity, the Peukert constant of the atomtronic battery is found to be $\rho = \mathrm{log}(I_L)/\mathrm{log}(C_c t_d) = 2.1(3)$. This large value of $\rho$ reflects the substantial dissipation within the atomtronic battery reservoir as current flows into the load.

The energy capacity $C_E = \int^{t_d}_0 I_L V_T dt$ of the atomtronic battery is an extension of $C_c$ and describes the total energy supplied by the battery. Figure~\ref{fig:BCc}(b) shows $C_E$, which is calculated by integrating the power output to the load during the discharge time.  Like the current capacity, the energy capacity shows an increasing trend with respect to the truncation parameter, in part due to the nearly linear increase in the terminal voltage. Ultimately, the energy capacity is limited by the total energy of the ensemble initially stored in the reservoir before discharge. For the data analyzed here this limit is $E_{tot} = 6.5(9)\times10^{5}$~kHz.

Lastly, the peak power output $P_{L,peak}$ is determined from experimental data using the relation $P_{L} = I_L V_T$, and is shown in Fig.~\ref{fig:BCc}(b). Output power of the atomtronic battery varies during the discharge time, providing a peak power immediately after connecting the battery to the load. The battery is capable of supplying a maximum peak power of $P_{L} \approx 5 \times 10^{4}$ kHz$^2$ for $\bar{\eta} < 3$, which correlates well with the regime of low internal resistance (see Fig.~\ref{fig:VTRint}(a)) as expected.  In spite of the increasing $V_{T}$, at higher truncation parameters the peak power output declines due to the exponentially decreasing $I_L$. Figure~\ref{fig:BCc} encapsulates the performance of the atomtronic battery and illustrates that the battery can be operated in regimes of either large peak output power (small $\bar{\eta}$) or large energy capacity (large $\bar{\eta}$).

As mentioned in Sec.~\ref{sec:Circuit}, a large energy capacity is desirable if long term, stable circuit operation is required because the battery discharges slowly in this regime, and maintains a stable terminal voltage for a longer period of time. Alternatively, operating with a large peak output power results in rapid discharge of the battery, but more instantaneous power available to drive a load. Typically, electronic batteries are used in the large energy capacity regime, and it is likely that this is where an atomtronic battery would also be operated. In this regime, the battery is capable of supplying a flux of atoms at a well defined energy for some period of time, which in principle allows one to study steady-state behavior of an atomtronic circuit.

\section{\label{sec:Conclusion}Conclusions\protect}
This work has demonstrated an atomtronic battery that supplies a thermal atom current to the connected load potential.  By monitoring the time evolution of the finite-temperature BEC that acts as the stored energy of the battery, all pertinent information regarding battery operation is extracted.  We model the discharge process with a set of rate equations, and the results are used to characterize the battery in terms of a Th\'{e}venin equivalent circuit containing a finite internal resistance. The internal resistance is found to be negative, which results in a decrease in temperature and sustained chemical potential as the battery supplies current to the load. This behavior is complementary to that of the atomtronic battery described in Ref.~\cite{2013:Zozulya} in which condensed atoms are supplied to the load resulting in a positive internal resistance and increase in temperature. 

In analyzing the characteristics of the atomtronic battery we compared the results to the behavior of electronic batteries. The chemical potential stored in the reservoir well was shown to remain nearly constant throughout the discharge, which reflects its role as the analogue to the electromotive force. Binary collisions attempt to maintain local thermal equilibrium in the reservoir well and repopulate atoms at the terminal voltage, which leads to sustained current output.  This mechanism is akin to the various reactions in electronic batteries that promote electrons to the proper electric potential at the battery terminals. As the battery discharges the terminal voltage decays in relation to the increasing magnitude of the internal resistance.

\ack
We thank A. A. Zozulya for useful discussions. This work was supported by the Air Force Office of Scientific Research (FA9550-14-1-0327) and the National Science Foundation (PHY1125844).

\end{document}